\documentclass[aps,prb,superscriptaddress,twocolumn,nopacs,amsmath,amssymb,letter]{revtex4}

\usepackage{color}
\usepackage{graphicx}
\usepackage{dcolumn}
\usepackage{bm}
\setcitestyle{numbers,square}

\begin{document}

\title{Narrow-band anisotropic electronic structure of ReS$_2$}
\author{D.~Biswas}
\affiliation {SUPA, School of Physics and Astronomy, University of St. Andrews, St. Andrews KY16 9SS, United Kingdom}
\author{Alex~M.~Ganose}
\affiliation{Department of Chemistry, University College London, 20 Gordon Street, London WC1H 0AJ, UK}
\affiliation{Diamond Light Source, Harwell Campus, Didcot, OX11 0DE, United Kingdom}
\author{R.~Yano}
\affiliation {Materials and Structures Laboratory, Tokyo Institute of Technology, Kanagawa 226-8503, Japan}
\author{J.~M.~Riley}
\affiliation {SUPA, School of Physics and Astronomy, University of St. Andrews, St. Andrews KY16 9SS, United Kingdom}
\affiliation{Diamond Light Source, Harwell Campus, Didcot, OX11 0DE, United Kingdom}
\author{L.~Bawden}
\author{O.~J.~Clark}
\author{J.~Feng}
\author{L.~Collins-Mcintyre}
\affiliation {SUPA, School of Physics and Astronomy, University of St. Andrews, St. Andrews KY16 9SS, United Kingdom}
\author{W.~Meevasana}
\affiliation {School of Physics, Suranaree University of Technology, Nakhon Ratchasima, 30000, Thailand}
\author{T.~K.~Kim}
\author{M.~Hoesch}
\affiliation{Diamond Light Source, Harwell Campus, Didcot, OX11 0DE, United Kingdom}
\author{J.~E.~Rault}
\affiliation{Synchrotron SOLEIL, CNRS-CEA, L'Orme des Merisiers, Saint-Aubin-BP48, 91192 Gif-sur-Yvette, France}
\author{T.~Sasagawa}
\affiliation {Materials and Structures Laboratory, Tokyo Institute of Technology, Kanagawa 226-8503, Japan}
\author{David~O.~Scanlon}
\affiliation{Department of Chemistry, University College London, 20 Gordon Street, London WC1H 0AJ, UK}
\affiliation{Diamond Light Source, Harwell Campus, Didcot, OX11 0DE, United Kingdom}
\author{P.~D.~C.~King}
\email{philip.king@st-andrews.ac.uk}
\affiliation {SUPA, School of Physics and Astronomy, University of St. Andrews, St. Andrews KY16 9SS, United Kingdom}

\begin{abstract}
We have used angle resolved photoemission spectroscopy to investigate the band structure of ReS$_2$, a transition-metal dichalcogenide semiconductor with a distorted 1T crystal structure. We find a large number of narrow valence bands, which we attribute to the combined influence of the structural distortion and spin-orbit coupling. We further image how this leads to a strong in-plane anisotropy of the electronic structure, with quasi-one-dimensional bands reflecting predominant hopping along zig-zag Re chains. We find that this does not persist up to the top of the valence band, where a more three-dimensional character is recovered with the fundamental band gap located away from the Brillouin zone centre along $k_z$. These experiments are in good agreement with our density-functional theory calculations, shedding new light on the bulk electronic structure of ReS$_2$, and how it can be expected to evolve when thinned to a single layer. 
\end{abstract}

\date{\today}

\maketitle 

\section{Introduction}
\label{sec:intro}

The semiconducting transition-metal dichalcogenides (TMDCs) with formula MX$_2$ (M$=$transition metal, X$=$chalcogen) have recently attracted much attention for their intriguing optical and electronic properties \cite{Wang2012_TMDC}. These are typically found to depend sensitively on material thickness, where thinning down prototypical compounds such as MoS$_2$ to a single layer drives a crossover from an indirect to a direct band gap semiconductor, accompanied with a dramatic increase in the photoluminescence yield \cite{Mak2010_ML_MoS2,Lee2010_MoS2Thickness, Sahin2013_WSe2Thickness}. In contrast, recent studies of the group VII TMDCs ReX$_2$ (X=Se, S) revealed a striking resilience of their measured photoluminescence to changing material thickness \cite{Tongay2014_decoupling,Zhao2015_LDep_ReSe2}, attributed to a particularly-weak interlayer coupling leading to a monolayer-like electronic structure in the bulk.  This opens the exciting prospect to achieve optoelectronic functionality from bulk ReX$_2$ of the form which can only be realised by complex fabrication of single-layer samples and devices in group VI semiconducting TMDCs. Coupled with a pronounced anisotropy in their measured optical and electrical properties \cite{Friemelt1993_PhotocurrentAni, Ho2007_DichoricOptElec, Chenet2015_RamanAni,Tiong1999_HallAni}, ReX$_2$ materials are therefore important compounds for expanding the functionality of the TMDCs class, and hold potential for next-generation technologies \cite{Liu2015_FET,Gao2016_storage}. The electronic structure underpinning their striking optoelectronic properties, however, remains almost completely unexplored experimentally to date. 

\begin{figure}[!b]
\center
\includegraphics[scale=.88]{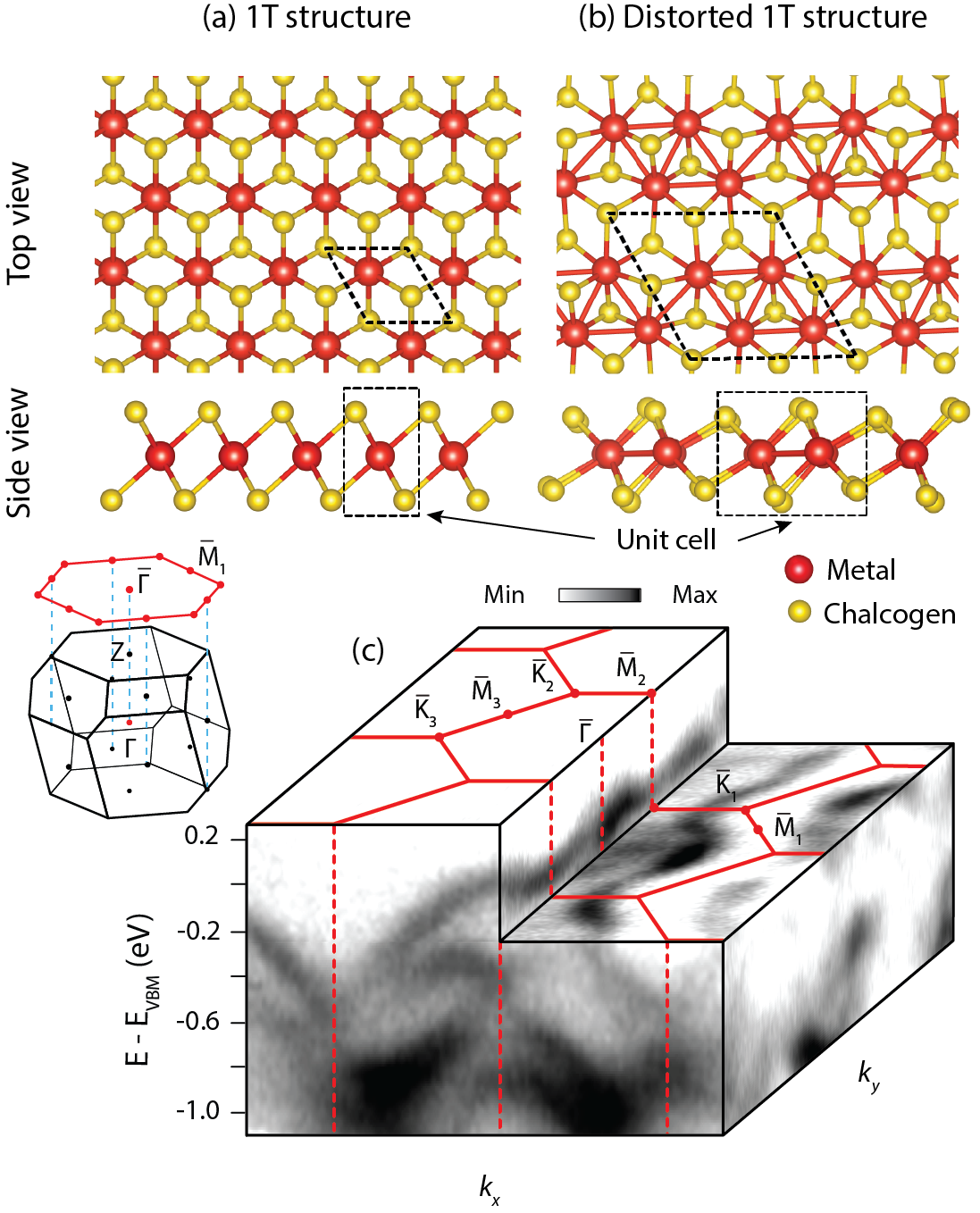} 
\caption{(a,b) Top and side views of a single layer (one unit cell in height) of an (a) un-distorted and (b) distorted 1T crystal structure, respectively. (c) Overview of the valence band structure as measured by ARPES, showing strong in-plane anisotropy. The surface Brillouin zone is shown as red lines. The bulk and projected surface Brillouin zones are shown in the inset.}
\label{fig:1}
\end{figure}

The crystal structure of ReS$_2$ (space group: P$\overline{\mathrm{1}}$) is shown in Fig.~\ref{fig:1}.  This can be described via a structural distortion away from the 1T structure that is found for some group V TMDCs such as TaS$_2$ and group X TMDCs such as PdTe$_2$. In the undistorted structure, the transition-metal sits at the centre of edge-sharing octahedra formed by the chalcogen atoms (Fig.~\ref{fig:1}(a)), with trigonal anti-prismatic point group symmetry of the transition metal site \cite{Chhowalla2013_TMDC}.  A Jahn-Teller-like structural distortion causes the formation of zig-zag Re chain-like structures as well as a pronounced out-of-plane buckling of the chalcogen layers (Fig.~\ref{fig:1}(b)) \cite{Kertesz1984_OctaVsTP,Tongay2014_decoupling}. The unit cell is approximately doubled in the in-plane directions, and the three-fold rotational symmetry of the parent 1T structure is lost. While the surface Brillouin zone remains almost hexagonal, here we show how the electronic structure becomes strongly anisotropic along different in-plane directions, with stripe-like band contours observed in angle-resolved photoemission (ARPES) measurements of the valence band constant energy contours (Fig.~\ref{fig:1}(c)). We will discuss this anisotropy in detail below, and show how it exhibits a striking binding-energy dependence tied to the underlying orbital character of the electronic states.

\section{Methods}
\label{sec:expt}
Single crystals of ReS$_2$ were grown by the chemical vapour transport technique with iodine as a transport agent. First, polycrystalline ReS$_2$ was prepared from a mixture of Re and S elements by the solid state reaction at 900$^{\circ}$C for 30 hours. The obtained ReS$_2$ powder and iodine were sealed in an evacuated quartz tube, and were heated using a two-heating zone furnace (1050$^{\circ}$C/950$^{\circ}$C) for 150 hours. Finally, the obtained crystals were cleaned by rinsing with acetone. This resulted in thin and flat crystals with reflective surfaces.  They were found to be insulating from resistivity measurements, while optical transmission measurements indicates a pronounced increase in optical absorption at an energy of $1.47\pm0.03$~eV, consistent with previous studies of the onset of direct optical absorption in this material \cite{Friemelt1996_OptBandGap,Ho1997_Bandedge_ReS2}.
 
ARPES measurements were performed using the I05 beamline of Diamond Light Source, UK and the CASSIOPEE beamline of SOLEIL synchrotron, France. Samples were cleaved {\it in situ} at the measurement temperature of between 10 and 20 K, and all measurements were performed using Scienta R4000 hemispherical electron analysers. From our photon energy-dependent measurements (Fig.~\ref{fig:3}), we estimate an inner potential $V_0=12$~eV employing a free-electron final state model, similar to the value found in other TMDCs \cite{Augustin2000_WTe2, Bovet2003_TaS2, Riley2014_SP,Bawden2016}. Due to the insulating nature of our samples, charging was evident as a shift of the spectral features to lower kinetic energy. We therefore reference our spectra to the valence band maximum (VBM), and compensate for relative charging shifts between spectra (e.g. when changing photon energy) by the relative shift of the Re 4{\it f} core level peak.  Surface doping was achieved by evaporating Rb on the sample at the measurement temperature from a well-outgassed SAES alkali metal source. We predominantly describe the electronic structure as measured within the almost hexagonal surface Brillouin zone, although the structural distortion causes neighbouring corners (face centres) of the surface Brillouin zone to become inequivalent. We thus label these with the conventional $\mathrm{\overline{K}}$ ($\mathrm{\overline{M}}$) notation, with an additional subscript to distinguish inequivalent symmetry points. The Re-Re zig-zag chains are oriented perpendicular to the $\overline{\Gamma}-\mathrm{\overline{M}_2}$ direction in our nomenclature.

All calculations were performed using the Vienna \textit{ab initio} Simulation Package (VASP) \cite{Kresse1993,Kresse1994,Kresse1996a,Kresse1996}, a periodic plane wave DFT code. The Projector Augmented Wave method was used to describe the interactions between core and valence electrons \cite{Kresse1999}. Convergence with respect to the plane wave basis set and \textit{k}-point sampling was performed, with a cut-off energy of 500$\,\mathrm{eV}$ and \textit{k}-point grid of $\Gamma$-centred $4 \times 4 \times 4$ found to be sufficient for the 12 atom unit cell of ReS$_2$. Geometry optimisations were performed using the PBEsol functional \cite{Perdew2008}, a version of the Perdew Burke and Ernzerhof (PBE) functional \cite{Perdew1996} revised for solids. PBEsol has previously been shown to adequately account for weakly dispersive interactions, such as those seen in the layered structure of ReS$_2$, without the need for an additional correction \cite{Travis2016,Ganose2016}. Optimisations were deemed converged when the sum of all forces on each atom totalled less than 10$\,\mathrm{meV}\,\textup{\AA}^{-1}$. The calculated structure parameters ($\mathrm{a} = 6.424\,\textup{\AA}, \mathrm{b} = 6.490\,\textup{\AA}, \mathrm{c}=6.407\,\textup{\AA},\alpha = 106.4^{\circ},\beta=88.2^{\circ},\gamma=121.4^{\circ}$) show good agreement with experiment (all within 0.6\%). In order to provide an accurate description of the electronic structure of ReS$_2$, the hybdrid functional, HSE06 \cite{Krukau2006}, was employed for band structure and density of states calculations. HSE06 combines 75\% exchange and 100\% of the correlation energies from PBE together with 25\% exact Hartree-Fock (HF) exchange at short ranges and has been shown to perform well for a wide range of solid-state semiconductors \cite{Maughan2016,Savory2016}. Special attention was paid to accurately modelling the relativistic effects seen in Re through use of scalar relativistic PAW pseudopotentials and explicit treatment spin--orbit coupling effects \cite{Hobbs2000}.

\begin{figure*}[!t]
\includegraphics[scale=.6]{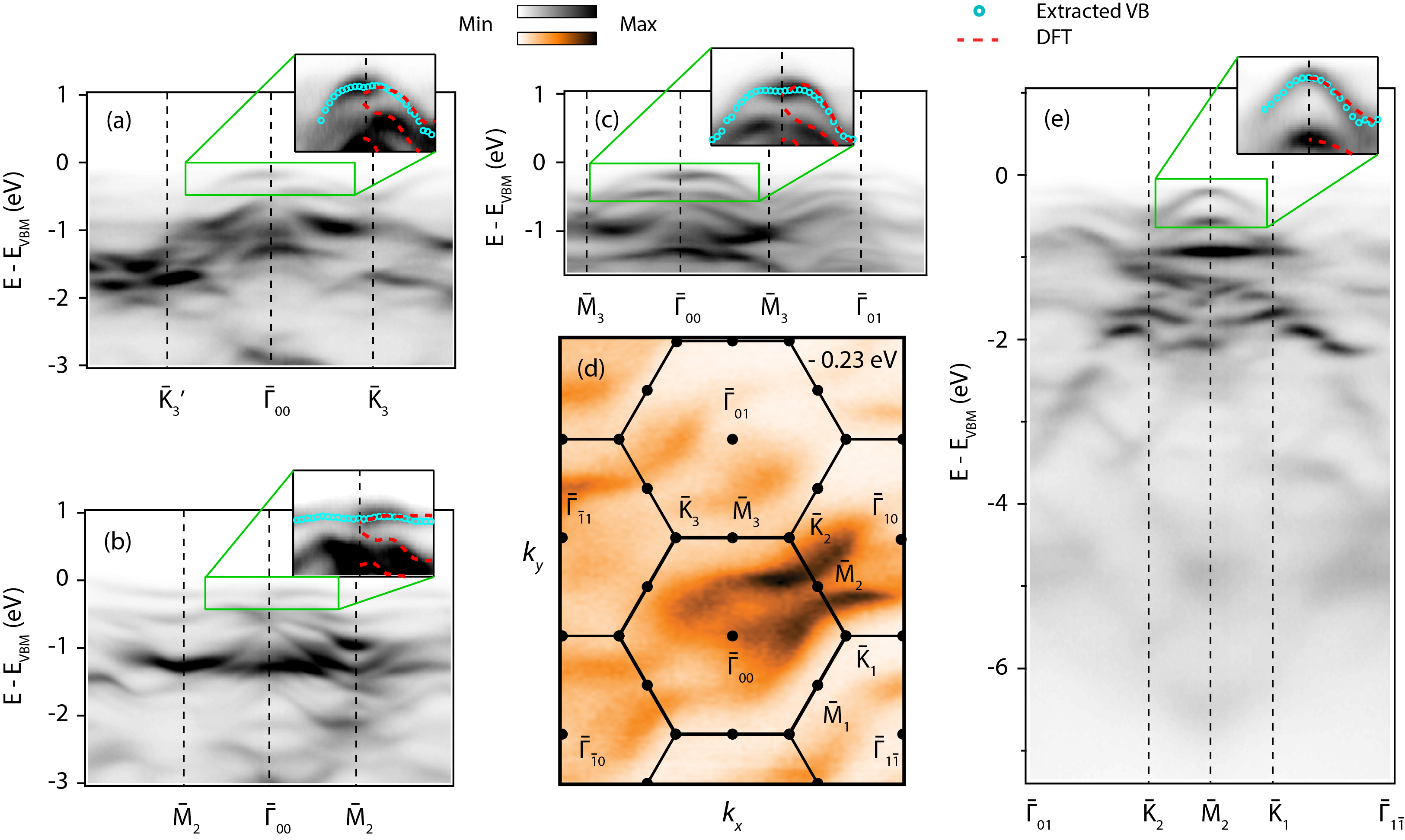} 
\caption{ARPES measurements ($h\nu=86$~eV, chosen to probe a bulk $\Gamma$-point along $k_z$) measured along the (a) $\overline{\Gamma}-\mathrm{\overline{K}_3}$, (b) $\overline{\Gamma}-\mathrm{\overline{M}_2}$, and (c) $\overline{\Gamma}-\mathrm{\overline{M}_3}$ directions of the surface Brillouin zone. These cut at different directions to the elongated contours visible in constant energy contours shown in (d) 230~meV below the VBM. The high-symmetry points are labelled in the panel. (e) Equivalent ARPES dispersions, shown over an extended binding energy range, measured along the boundary of the surface Brillouin zone ($\mathrm{\overline{K}_2-\overline{M}_2-\mathrm{\overline{K}_1}}$). For all measurements, the detailed dispersion of the uppermost valence band is shown inset, extracted from fits to EDCs and shown together with corresponding calculated valence band dispersions from density-functional theory.}
\label{fig:2}
\end{figure*}

\section{Results and discussions}
\label{sec:results}
We start by considering the detailed valence band dispersions measured along different high-symmetry directions of the surface Brillouin zone (Fig.~\ref{fig:2}). At the photon energy used, these measurements probe a $\Gamma$-point of the bulk Brillouin zone for $k_\parallel=0$, although we note that we have a finite $k_z$-resolution which we estimate as 0.17~\AA$^{-1}$  due to the surface sensitivity of photoemission, while the value of $k_z$ probed also decreases with increasing in-plane momentum \cite{Stroco2003_IntrinsicBroadening_PES}. For the measurements shown here, the value of $k_z$ typically varies by less than 10\% of the Brillouin zone height with varying in-plane momentum across the first surface Brillouin zone. 

\begin{figure*}[!t]
\includegraphics[scale=.72]{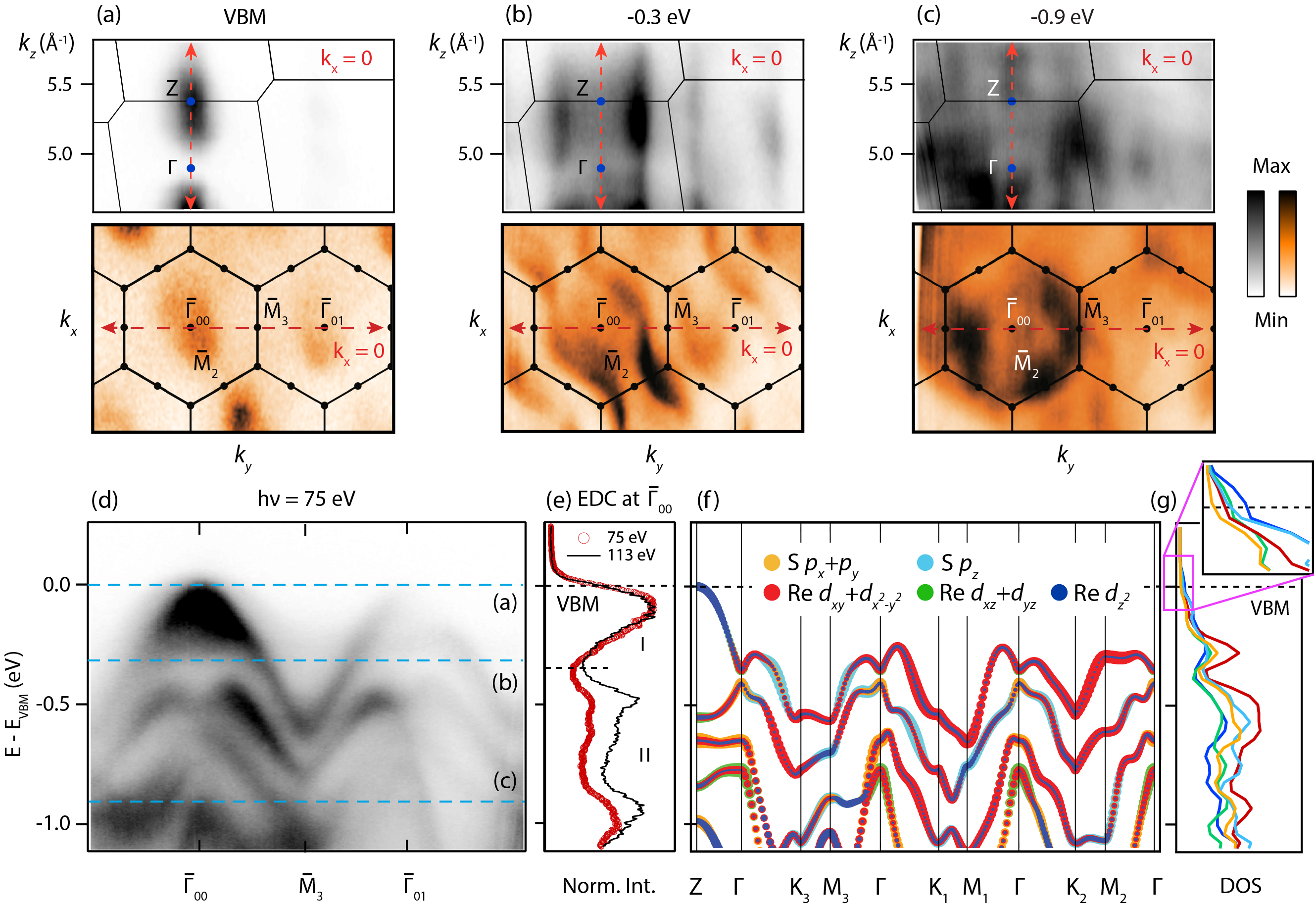} 
\caption{(a-c) $k_z$ dependence of the MDC at $k_x=0$ (top) and in-plane constant energy contours at a photon energy $h\nu=86$~eV, corresponding to $k_z\sim\!10\pi/c$, (a) at the valence band top and (b) 0.3~eV and (c) 0.9~eV below the VBM respectively, as shown by the blue lines in the dispersion $(h\nu=75$~eV) shown in (d). (e) Comparison of energy dispersive curves (EDCs) at $k_\parallel=0$ taken using 75 eV and 113 eV photons, showing an enhanced Re-derived orbital character closer to the band top. (f - g) Corresponding orbitally-resolved calculations from density-functional theory.}
\label{fig:3}
\end{figure*}

It is immediately evident that there are a large number of bands within $\sim\!3$~eV of the valence band top, each with a narrow bandwidth typically on the order of only 200~meV. This initially appears counterintuitive, given the spatially-extended nature of the Re $5d$ and S $3p$ orbitals from which these valence states derive.  However, doubling of the unit cell in both in-plane directions would cause a back-folding of the broad bands that would be expected in the undistorted 1T crystal structure. This will drive multiple band crossings throughout the Brillouin zone. Strong spin-orbit coupling can be expected to open pronounced hybridisation gaps at these band crossings, thereby creating much narrower bandwidths than would naively be expected considering the undistorted crystal structure. Indeed, tracing the dominant spectral weight, which normally follows the unreconstructed band dispersion in ARPES spectra \cite{Ku2010_bandUnfolding, Lee2010_BandFolding_ARPES}, reveals much broader bandwidths from which these narrow bands ultimately result. This is particularly evident along the $\mathrm{\overline\Gamma_{01}-\overline\Gamma_{1\bar{1}}}$ direction (Fig.~\ref{fig:2}(e)), which would be a $\mathrm{\overline{K}-\overline{K}}$ direction in the un-reconstructed Brillouin zone, where broad band-like variations in the spectral weight can be traced dispersing over several electron volts width.  We thus attribute the presence of multiple bands with narrow bandwidths here to the combined effects of structural distortion and strong spin-orbit coupling characteristic of the Re $d$-orbital manifold. A similar mechanism for narrow-band formation was recently proposed in the strong spin-orbit semimetal SrIrO$_3$ \cite{Nie_2015_SOCinSrIrO3}.

Even within these narrow bands, we find significant differences in bandwidth along different momentum directions. In Fig.~\ref{fig:2}(a-c), we show the valence band dispersions measured along $ \mathrm{\overline{\Gamma} - \overline{K}_3}$, $\mathrm{\overline{\Gamma} - \overline{M}_2}$, and $ \mathrm{\overline{\Gamma} - \overline{M}_3}$ of the surface Brillouin zone, respectively. Focusing on the uppermost valence band, we find the most dispersive states along $ \mathrm{\overline{\Gamma} - \overline{K}_3}$ (and parallel to this, along $ \mathrm{\overline{K}_2 - \overline{M}_2 - \overline{K}_1} $, Fig.~\ref{fig:2}(e)). This corresponds to the direction along the Re chains formed by the structural distortion (Fig.~\ref{fig:1}(b)). In a tight-binding picture, hopping should therefore be maximal along this direction, leading to the most dispersive states. Even along this direction, however, the valence band at $\Gamma$ is not parabolic, instead flattening over to form a slightly ``M''-shaped dispersion. This is evident in fits to our measured energy-distribution curves (EDCs), as well as our density-functional theory calculations which are in good agreement with our measured dispersions (Fig.~\ref{fig:2}(a), inset). This will lead to an enhanced density of states near the band edge, helping to explain the strong optical absorption and photoluminescence which this compound is known to possess \cite{Tongay2014_decoupling}. On moving away from this direction, this band top gradually flattens (Fig.~\ref{fig:2}(c)), before becoming almost non-dispersive along the $\mathrm{\overline{\Gamma} -\overline{M}_2 }$ direction (normal to the chain). This points to a limited hopping between the chains, and thus a highly-anisotropic in-plane electronic structure. This is also evident from a constant energy contour shown 230~meV below the valence band top in Fig.~\ref{fig:2}(d), which exhibits band contours that are closed along the $\mathrm{\overline{\Gamma} -\overline{K}_3} $ direction, but open along the orthogonal $\mathrm{\overline{\Gamma} -\overline{M}_2 }$ direction. This directly underpins the strong anisotropy of in-plane transport of this compound \cite{Ho2007_DichoricOptElec}.

It is thought that the coupling strength between the layers of ReS$_2$ is very weak \cite{Tongay2014_decoupling}. The electronic states would naturally therefore be expected to be predominantly confined within single S-Re-S layers which, coupled with their strongly anisotropic in-plane dispersion, would result in quasi-one-dimensional electronic states. In contrast, our photon energy-dependent ARPES measurements (Fig.~\ref{fig:3}) indicate that this is only true within limited binding energy ranges. Fig.~\ref{fig:3}(a-c) shows the dispersion of electronic states at $k_x=0$ as a function of $k_z$ (top) and for varying in-plane momentum for approximately fixed $k_z=10\pi/c$ ($h\nu=86$~eV, bottom). These are shown at the valence band maximum (VBM, Fig.~\ref{fig:3}(a)), and 300 and 900~meV below the VBM, respectively (Fig.~\ref{fig:3}(b,c)). Focusing first at the VBM (Fig.~\ref{fig:3}(a)), rather than being quasi-one-dimensional, we find that the electronic bands are in fact quasi-three-dimensional. Elongated ellipses are found along both the out-of-plane and in-plane momentum directions, but both form closed contours located around the Z-point of the bulk Brillouin zone (BZ). We therefore find that the VBM is not located at the BZ centre, but rather at the BZ face along $k_z$.

In contrast, only 300~meV below the VBM, the valence band states exhibit only very little out-of-plane dispersion, evident from the almost invariant band momenta with varying photon energy (Fig.~\ref{fig:3}(b)). By this binding energy, they have also developed the open contours within the transition-metal plane as discussed above, and so their electronic states can best be described as quasi-one-dimensional in this energy range. At 900~meV below the VBM, the band contours reach close to the BZ boundary within the plane, and a complex band structure is evident along $k_z$, with several bands showing significant $k_z$ dispersion. Our measurements therefore reveal a rich binding energy dependence of the dimensionality of the electronic structure of ReS$_2$.

\begin{figure}[!t]
\includegraphics[scale=.53]{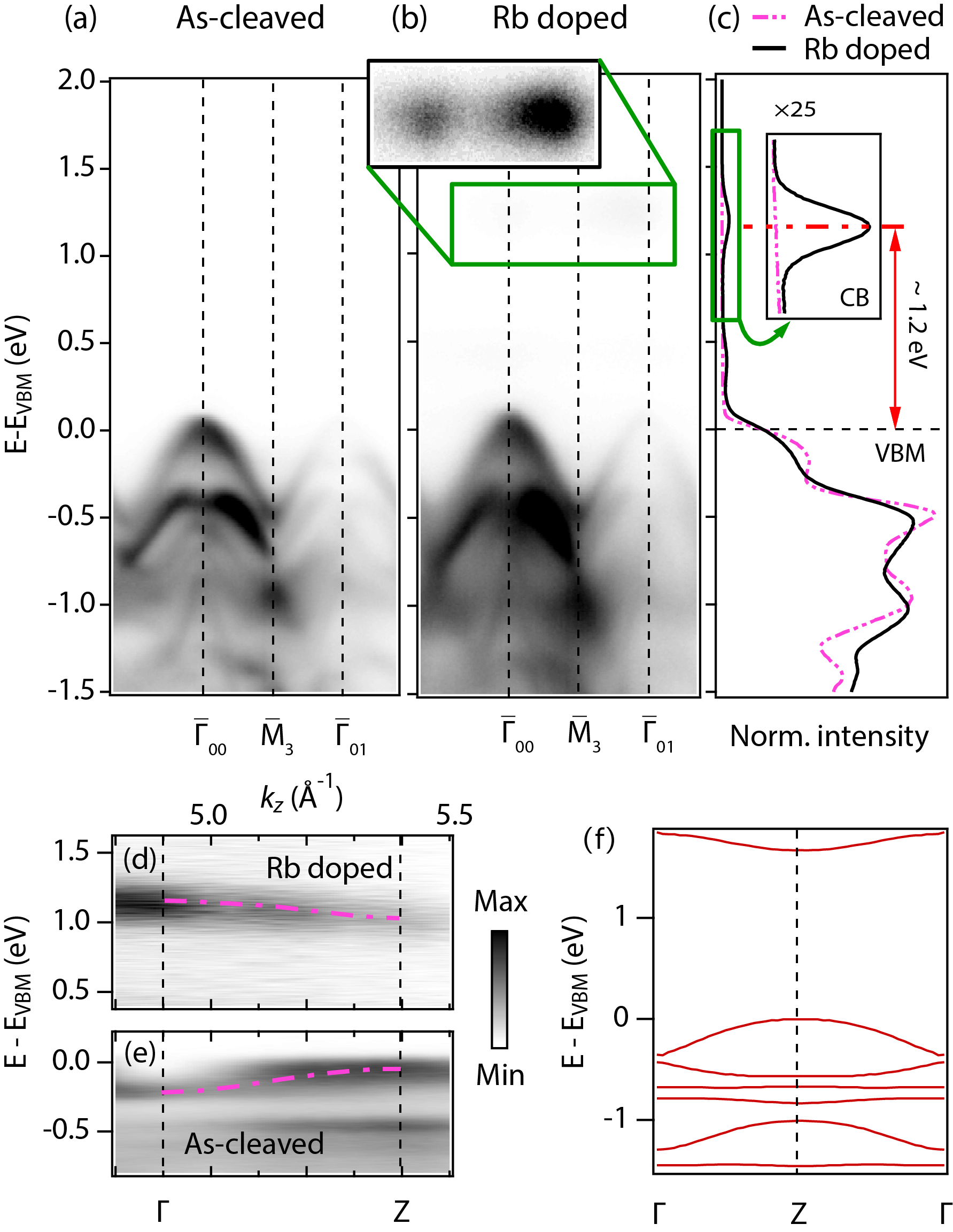} 
\caption{(a) and (b) ARPES spectra for as-cleaved and Rb doped ReS$_2$ along $\mathrm{\overline{\Gamma} - \overline{M}_3}$. The inset shows the energy region around the conduction band states in higher contrast. (c) Integrated spectra over two BZs for the as-cleaved and doped sample, again showing the filling of the conduction band. (d) and (e) $k_z$ dispersion measured at $k_\parallel=0$ for doped and clean ReS$_2$. (f) Calculated $k_z$ dispersion along $\mathrm{\Gamma - Z}$ from DFT.}
\label{fig:4}
\end{figure}

We show below how this arises due to a variation in orbital character of the valence bands. Fig.~\ref{fig:3}(f -g) show the orbitally-projected electronic structure calculated from DFT. While the orbital character is strongly mixed throughout the valence bands, our calculations indicate that out of plane Re $5d_{3z^2-r^2}$ orbitals contribute a significant weight close to the VBM. Such out-of-plane orbitals mediate larger inter-layer hopping, consistent with the quasi-3D nature of the electronic states at the VBM. In contrast, in-plane $d$-orbitals dominate slightly ($\gtrsim250$~meV) below the VBM, leading to the quasi-1D dispersions observed above. Deeper still, we find a more significant contribution of S $3p$ orbitals. Located on either side of the van der Waals gap, these naturally mediate a more significant inter-layer hopping than the Re-derived states, explaining the re-entrant three-dimensionality observed in our experiment. 

This is entirely consistent with our photon energy-dependent ARPES data. We show in Fig.~\ref{fig:3}(e) two EDCs measured at the $\overline{\Gamma}$ point using photon energies of  75 and 113~eV. These are selected to probe the same point in $k_z$ (the mid-point along $\mathrm{\Gamma -Z}$) and to lead to the greater difference in photoionization cross-section of Re $5d$ and S $3p$ states, with the former photon energy almost 3 times more sensitive to Re- vs.\ S-derived states than the latter \cite{YehLindau}. A significantly greater relative spectral weight more than $\sim\!350$~meV below the VBM for measurements at $h\nu=113$~eV points to a much larger S $3p$-derived character of the valence band states here than at the VBM, as found in our calculations. Together, these measurements and calculations reveal that the dimensionality of electronic states in ReS$_2$ is inherently tied to their varying orbital makeup.

A key question across the TMDC materials class has been the nature and dimensional-control of the fundamental band gap. To address this here, we have electron-doped the surface of ReS$_2$ via deposition of small concentrations of Rb atoms, in order to populate the conduction band states. Our measured ARPES spectra from as-cleaved and Rb-doped samples are shown in figure \ref{fig:4}(a) and (b), respectively. We find that new states are populated $\sim\!1.2$~eV above the VBM, located at the $\overline{\Gamma}$ point of the surface Brillouin zone. We thus attribute these as occupied conduction band states. Our photon energy-dependent ARPES data (Fig,~\ref{fig:4}(d,e)) indicate that not only the VBM, but also the conduction band minimum (CBM) have a significant dispersion along $k_z$. The conduction and valence bands disperse in opposite sense, ensuring a direct band gap with high joint density of states persists in bulk ReS$_2$, albeit located at the BZ boundary along $k_z$ rather than at the BZ centre. This is supported by our DFT calculations (Fig.~\ref{fig:4}(f)). While the out-of-plane dispersion is slightly overestimated by these calculations, suggesting too high an inter-layer interaction in our theory, they are in excellent qualitative agreement with our experimental measurements of a direct band gap at the $\mathrm{Z}$-point of the bulk Brillouin zone. Nonetheless, we note we cannot exclude a slightly smaller indirect band gap located away from high symmetry points, as suggested by recent optical measurements \cite{Aslan2016_Exciton_ReS2}.

Intriguingly, the band gap we extract from our ARPES measurements ($\sim\!1.2$~eV) is $\sim\!300$~meV smaller than the measured optical band gap. The surface doping approach used here could be expected to create a near-surface downward band bending, rather than inducing a rigid band filling of the conduction band states \cite{king2008_coupledPSsolution, king2008_ChargeNeutralityIn2O3, King2011_LargeTunableRashba_Bi2Se3, Bahramy2012_QuantumConfinement_TI}. This, however, would be expected to lead to an increased band gap via quantum confinement of the conduction band states, and is inconsistent with the three-dimensional dispersions of the conduction band states which we observe. The smaller band gap observed here may instead indicate an increased electronic screening due to the high near-surface electron density, creating a strong renormalisation of the electronic band gap from its value in the undoped semiconductor, as has been observed in other TMDC compounds \cite{Ugeda_2014_ExcitonicEffectTMDC, Riley2015_WSe2}. This possibility requires further exploration, both experimentally and theoretically, but may point to the presence of rather strongly bound excitons even in bulk ReS$_2$. This would seem broadly consistent with the observed pronounced excitonic features in optical spectra \cite{Tongay2014_decoupling, Aslan2016_Exciton_ReS2}.

\section{Conclusion}
\label{sec:conclu}
In conclusion, our study has revealed the bulk electronic structure of the TMDC semiconductor ReS$_2$ for the first time. We have imaged how the presence of an in-plane structural distortion leading to the formation of Re zig-zag chains, combined with weak interlayer coupling in this compound, drives the formation of quasi-one-dimensional electronic states in its valence bands, underpinning its anisotropic optoelectronic properties.. However, we have also shown how this breaks down both near to the band edges, and at higher binding energies, where a more significant three-dimensionally of the electronic states results due to the mixing in of atomic orbitals with out-of-plane character. The resulting electronic structure still leads to a direct or nearly-direct band gap semiconductor in bulk, unlike for other bulk TMDCs such as MoS$_2$, but with the band extrema located away from the zone centre along $k_z$. The $k_z$-dispersion of the band edge states naturally explains the blueshift observed in the peak energy of photoluminescence when ReS$_2$ is thinned to a single monolayer \cite{Tongay2014_decoupling}, as quantum confinement in the $z$-direction will increase the energy of the direct band gap in this compound. Together, this opens exciting new possibilities for tuning and utilising the anisotropic electrical and optical response of ReS$_2$ bulk and thin-film samples.

{\it Note: While finalising our work, a paper appeared on arXiv also reporting a non-negligible out-of-plane dispersion of ReS$_2$ \cite{Gehlmann2017_arxiv_ReS2}. Their data seems consistent with ours. We are also grateful to Dr.\ Daniel Wolverson, University of Bath, UK, for discussing the results of their similar measurements with us prior to publication, which also appear to be consistent with the data presented here. }

\section{Acknowledgement}
This work was supported by the Engineering and Physical Sciences Research Council, UK (Grant No.~EP/M023427/1).  This work also made use of the ARCHER UK National Supercomputing Service (http://www.archer.ac.uk), via membership of the UK's HEC Materials Chemistry Consortium, which is funded by EPSRC (EP/L000202). We thank Diamond Light Source (via Proposal Nos.~SI9500 and SI11383) and SOLEIL synchrotrons for access to Beamlines I05 and CASSIOPEE, respectively, that contributed to the results presented here.  AMG acknowledges Diamond Light Source for the co-sponsorship of a studentship on the EPSRC Centre for Doctoral Training in Molecular Modelling and Materials Science (EP/L015862/1). JMR, LB and OJC acknowledge EPSRC for PhD studentship funding through grant Nos.~EP/L505079/1, EP/G03673X/1, and EP/K503162/1, respectively. WM got support from Thailand Research Fund and Suranaree University of Technology (Grant No.~BRG5880010). TS was supported by a CREST project from Japan Science and Technology Agency (JST) and a Grants-in-Aid for Scientific Research (B) [16H03847] from Japan Society for the Promotion of Science (JSPS). DOS acknowledges support from the EPSRC (EP/N01572X/1). DOS and PDCK acknowledge membership of the Materials Design Network.

\end{document}